\documentclass[reprint,amsmath,amssymb,longbibliographyaps,prd]{revtex4-2}

\usepackage{graphicx}
\usepackage[colorinlistoftodos]{todonotes}
\usepackage{siunitx}
\usepackage{booktabs}
\usepackage{xcolor}

\usepackage{hyperref}
\hypersetup{
    colorlinks=true,
    linkcolor=red,
    anchorcolor=black,
    citecolor=blue,
    filecolor=cyan,
    menucolor=red,
    runcolor=cyan,
    urlcolor=magenta,
    pdftitle={Cross-Polarized Stimulated Brillouin Scattering in Lithium Niobate Waveguides},
    pdfpagemode=FullScreen,
    }
\usepackage{cleveref}
\Crefname{equation}{Eq.}{Eqs.}
\Crefname{figure}{Fig.}{Figs.}
\Crefname{table}{Tab.}{Tabs.}

\begin{document}

\title{Cross-Polarized Stimulated Brillouin Scattering in Lithium Niobate Waveguides}

\author{Caique C. Rodrigues$^{1,2}$}
\affiliation{\vspace{0.25cm} $^{1}$Gleb Wataghin Physics Institute, University of Campinas, Campinas, SP, Brazil \\
$^{2}$John A. Paulson School of Engineering and Applied Sciences, Harvard University, Cambridge, Massachusetts 02138, USA}

\author{Nick J. Schilder$^{1}$}
\affiliation{\vspace{0.25cm} $^{1}$Gleb Wataghin Physics Institute, University of Campinas, Campinas, SP, Brazil \\
$^{2}$John A. Paulson School of Engineering and Applied Sciences, Harvard University, Cambridge, Massachusetts 02138, USA}

\author{Roberto O. Zurita$^{1}$}
\affiliation{\vspace{0.25cm} $^{1}$Gleb Wataghin Physics Institute, University of Campinas, Campinas, SP, Brazil \\
$^{2}$John A. Paulson School of Engineering and Applied Sciences, Harvard University, Cambridge, Massachusetts 02138, USA}

\author{Letícia S. Magalhães$^{2}$}
\affiliation{\vspace{0.25cm} $^{1}$Gleb Wataghin Physics Institute, University of Campinas, Campinas, SP, Brazil \\
$^{2}$John A. Paulson School of Engineering and Applied Sciences, Harvard University, Cambridge, Massachusetts 02138, USA}

\author{Amirhassan Shams-Ansari$^{2}$}
\affiliation{\vspace{0.25cm} $^{1}$Gleb Wataghin Physics Institute, University of Campinas, Campinas, SP, Brazil \\
$^{2}$John A. Paulson School of Engineering and Applied Sciences, Harvard University, Cambridge, Massachusetts 02138, USA}

\author{Felipe J. L. dos Santos$^{1}$}
\affiliation{\vspace{0.25cm} $^{1}$Gleb Wataghin Physics Institute, University of Campinas, Campinas, SP, Brazil \\
$^{2}$John A. Paulson School of Engineering and Applied Sciences, Harvard University, Cambridge, Massachusetts 02138, USA}

\author{Otávio M. Paiano$^{1}$}
\affiliation{\vspace{0.25cm} $^{1}$Gleb Wataghin Physics Institute, University of Campinas, Campinas, SP, Brazil \\
$^{2}$John A. Paulson School of Engineering and Applied Sciences, Harvard University, Cambridge, Massachusetts 02138, USA}

\author{Thiago P. M. Alegre$^{1}$}
\affiliation{\vspace{0.25cm} $^{1}$Gleb Wataghin Physics Institute, University of Campinas, Campinas, SP, Brazil \\
$^{2}$John A. Paulson School of Engineering and Applied Sciences, Harvard University, Cambridge, Massachusetts 02138, USA}

\author{Marko Lončar$^{2}$}
\affiliation{\vspace{0.25cm} $^{1}$Gleb Wataghin Physics Institute, University of Campinas, Campinas, SP, Brazil \\
$^{2}$John A. Paulson School of Engineering and Applied Sciences, Harvard University, Cambridge, Massachusetts 02138, USA}

\author{Gustavo S. Wiederhecker$^{1}$}
\affiliation{\vspace{0.25cm} $^{1}$Gleb Wataghin Physics Institute, University of Campinas, Campinas, SP, Brazil \\
$^{2}$John A. Paulson School of Engineering and Applied Sciences, Harvard University, Cambridge, Massachusetts 02138, USA}

\begin{abstract}
    We report on the experimental demonstration of cross-polarization backward stimulated Brillouin scattering (BSBS) in lithium niobate on insulator (LNOI) waveguides. Performing polarization-sensitive pump and probe measurements, we captured both intra- and intermodal scattering between counterpropagating fundamental optical modes. Remarkably, cross-polarization scattering achieved SBS gains that exceeded $G_{B}=\SI{80}{\m^{-1}\W^{-1}}$. This substantial gain not only broadens the utility of polarization in SBS but also paves the way for high-performance devices, including ultranarrowband lasers, robust broadband nonreciprocal devices, RF filters, and microwave-to-optical converters.
\end{abstract}

\maketitle

Lithium niobate (LN) has been essential in bulk optics and telecommunication, prized for its electro-optic and nonlinear optical properties. With integrated photonics on the rise, the ability of LN to tightly confine optical and acoustic waves has gained attention to enhance wave interactions~\cite{Zhu:21,Jiang:19,Sarabalis:21,Jiang2016,Shao:19}. Despite its intrinsic material properties, achieving the simultaneous confinement of both optical and mechanical waves in LN-based integrated photonics requires careful design considerations. Current strategies include fully suspended structures~\cite{VanLaer2015} and the excitation of surface acoustic waves (SAW) through interdigital transducers (IDT)~\cite{Liu:19}. Backward stimulated Brillouin scattering (BSBS) has been explored as a means of excitation of short-wavelength traveling acoustic waves at several GHz frequencies and beyond~\cite{Wolff:21,10.1063/1.5088169,10.5555/1817101,Eggleton2019,Eggleton:13,PhysRevX.2.011008,PhysRevA.92.013836}. Previous SBS studies in integrated photonics have focused on silicon~\cite{Shin2013,Kittlaus2017,ye2024stimulatedbrillouinscatteringnonsuspended,dinter2024antiresonantreflectingacousticrib}, silicon nitride~\cite{doi:10.1126/sciadv.abq2196,PhysRevLett.124.013902,botter2023stimulated,klaver2024surfaceacousticwavesbrillouin}, chalcogenide glasses~\cite{Pant:11,Kabakova:13,Du:16}, and doped silica chip waveguides~\cite{ZERBIB2023106830}, enabling demonstrations of high spectral purity Brillouin lasers~\cite{Gundavarapu2019, Chauhan2021,Li:14,8925242}, nonreciprocal optoacoustics~\cite{Sohn2019,Lai:23,Poulton:12,PhysRevX.14.021002,https://doi.org/10.1002/lpor.201900278}, radio-frequency (RF) photonics~\cite{10.1063/1.5113569}, light storage systems~\cite{10.1063/5.0193174,Stiller:20,Stiller2024}, and quantum-enabled functionalities~\cite{PhysRevB.110.014416,diamandi2024quantumoptomechanicalcontrollonglived,Li:22,PhysRevResearch.5.043140,zhu2024optoacousticentanglementcontinuousbrillouinactive,PhysRevLett.132.023603}. Recently, the first on-chip SAW SBS using a GeAsSe platform was observed~\cite{10.1063/5.0220496}, and demonstrations of on-chip SBS in lithium niobate on insulator (LNOI) have also emerged~\cite{ye2023surface,s11433-023-2272-y,yu2024onchipbrillouinamplifiersuspended}. Despite progress, controlling the optical polarization state remains a challenge due to the longitudinal nature of acoustic waves in BSBS, which limits the interaction between orthogonal polarizations~\cite{PhysRevB.19.4986,285349}. In contrast, using forward SBS (FSBS) in optical fibers, manipulation of optical polarization states has been demonstrated~\cite{10.1063/5.0040580,culverhouse,diamandi,Bashan2022}, at the cost of operating at lower mechanical frequencies.

Here we present the first demonstration of cross-polarized backward SBS within an integrated photonic structure. This demonstration unlocks the potential for exploring the vectorial properties of SBS explored in optical fibers~\cite{Zadok:08} and free-space cavities~\cite{nie2024crosspolarized} within integrated photonics platforms, enabling SBS-based polarization converters and filters. By examining different waveguide angles and widths, we improved the understanding of SBS in anisotropic materials and uncover the potential of on-chip polarization dependent SBS. Creating a frequency-selective Brillouin gain for orthogonally polarized light may help filter strong residual pump light in critical applications, such as high-purity SBS lasers or preparing mechanical quantum states with heralded single-photon detection~\cite{Vanner2015TowardsMotion,PhysRevLett.126.033601}.

\begin{figure}[ht]
    \centering
    \includegraphics{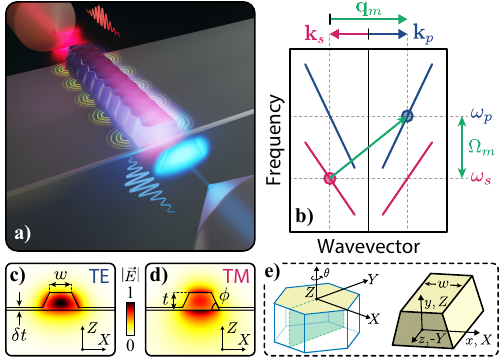}
    \caption{\textbf{(a)} Cross-polarized BSBS in an integrated waveguide. A pump and a probe beam travel in opposite directions with different polarizations, creating an interference pattern that triggers acoustic vibrations; \textbf{(b)} Illustrative conservation of energy and momentum diagram as a pump photon $\left(\omega_{p}, \mathbf{k}_{p}\right)$ splits into a backward-propagating photon $\left(\omega_{s}, \mathbf{k}_{s}\right)$ and a copropagating phonon $\left(\Omega_{m}, \mathbf{q}_{m}\right)$; \textbf{(c,d)} Fundamental TE and TM electric field profiles simulated using COMSOL multiphysics; \textbf{(e)} On the left, the LN unit cell is depicted with its crystallographic axes and one of the three possible mirror symmetry planes (shown in green). On the right, the relative orientation between the LN crystal and the waveguide is illustrated. Uppercase letters ($X, Y, Z$) denote the material frame, while lowercase letters ($x, y, z$) represent the lab frame. Throughout this text, all rotations are defined with respect to the material's $Z$-axis, indicated in the figure by $\theta$.}
    \label{fig:Fig1}
\end{figure}

\begin{figure*}[ht]
    \centering
    \includegraphics{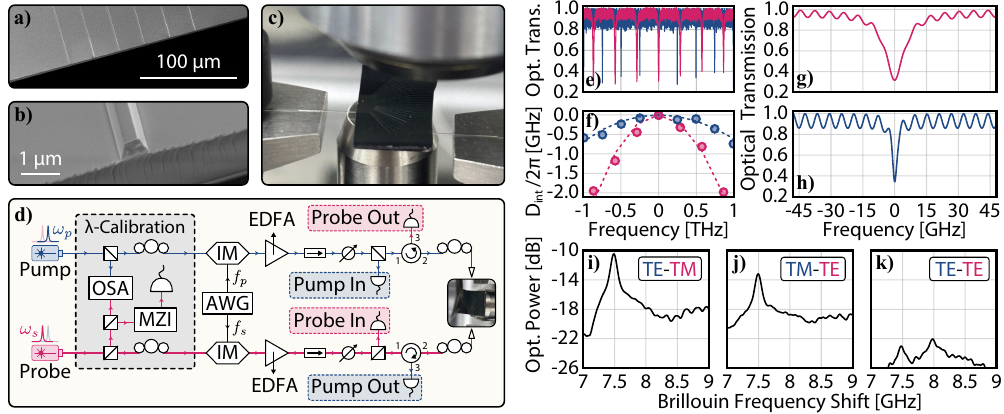}
    \caption{\textbf{(a,b)} Scanning electron microscope (SEM) images of a waveguide family; \textbf{(c)} Experimental setup showing coupling to the sample via lensed fibers; \textbf{(d)} Experimental setup schematic. IM: intensity modulator, AWG: arbitrary waveform generator, OSA: optical spectrum analyzer, MZI: Mach-Zehnder interferometer, EDFA: Erbium-doped fiber amplifier; \textbf{(e,f)} Optical spectrum of our \SI{80}{\um} radius LN ring resonator and its second order dispersion, respectively. The red curve corresponds to TM modes, while the blue curve corresponds to TE modes; \textbf{(g,h)} Central TM and TE modes, respectively; \textbf{(i,j,k)} Optical output power signal showing the polarization-dependent results for pump and probe combinations. The vertical scale is normalized by the peak Brillouin signal from the optical fiber patch cords, which is at 0 dB and not shown. The labels ``TE-TM" and similar denote the polarizations of the pump (blue) and probe (red) in each set of data, respectively. This data was measured using the waveguide with $w = $ \SI{500}{\nm} and $\theta = $ \SI{0}{\degree}.}
    \label{fig:Fig2}
\end{figure*}

The excitation of acoustic waves through SBS occurs due to optical forces induced by the intensity beating pattern caused by the interference of counterpropagating fields~\cite{10.1063/1.1654249}. Due to energy and momentum conservation, the resulting acoustic wavelength matches the spatial period of this beating pattern, which is half the optical wavelength~\cite{10.1063/1.5088169}. Depending on the contrast in the wave velocities, these acoustic waves can be partially guided in the substrate-anchored LNOI waveguides~\cite{Rodrigues:23}. We illustrate the concept of cross-polarized BSBS in \Cref{fig:Fig1}(a,b), where the optical pump wave (pump frequency $\omega_{p}$) and the red-shifted probe wave (Stokes frequency $\omega_{s}$) are launched in opposite directions in the LNOI waveguide into orthogonal transverse-electric (TE, \Cref{fig:Fig1}c) and transverse-magnetic (TM, \Cref{fig:Fig1}d) modes. When the optical frequencies and optical wavevectors of the pump and probe are separated by the acoustic frequency ($\Omega_{m}=\omega_{p}-\omega_{s}$) and by the acoustic wavevector ($\mathbf{q}_{m}= \mathbf{k}_{p} - \mathbf{k}_{s}$), a shear SAW can be resonantly excited, which induces optical gain by transferring energy from the pump to the probe wave.
 
The waveguides used in our demonstration are fabricated from a \SI{400}{\nm}-thick $Z$-cut LN layer on top of \SI{4.7}{\um} of silicon dioxide (SiO$_{2}$) grown on a silicon carrier wafer~(NanoLN)~\cite{Zhang:17,Li2023}. The LN thin film is only partially etched, leaving a residual thickness of \SI{30}{\nm}. In \Cref{fig:Fig2}(a,b) we show electron microscopy images of the waveguides, designed to exclusively support the propagation of the fundamental TE and TM optical modes. Due to the high anisotropy of LN, SBS is expected to have a nontrivial polarization dependence, although it has not been previously explored~\cite{Rodrigues:23,ye2023surface}. To investigate the anisotropic properties of the optomechanical interactions, the chip design features an array of waveguide sets, each set oriented at a specific angle $\theta$ (\Cref{fig:Fig1}e) relative to the material's $Y$-axis. A full description of the fabrication recipe and sample details can be found in \Cref{supp:Sample_Details_and_Fabrication_Recipe} of Supplementary Material.

\begin{figure*}[ht]
    \centering
    \includegraphics{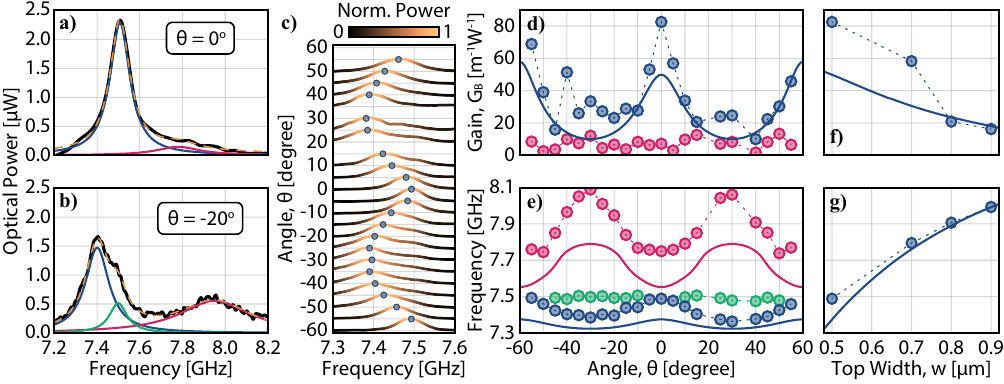}
    \caption{\textbf{(a,b)} Experimental BSBS TE-TM spectrum for $\theta = $ \SI{0}{\degree} and $\theta = $ \SI{-20}{\degree}, respectively, at fixed waveguide top width $w =$ \SI{500}{\nm}. The black curves are the experimental data, the yellow dashed curves are the fitting as we sum the colored Lorentzian modes; \textbf{(c)} Evolution of the Brillouin gain spectrum as we vary the orientation of the waveguide. The $y$-axis corresponds to the angle of each peak, represented in blue; \textbf{(d,e)} The dots are the experimentally measured Brillouin gain and peak frequency of the modes as function of the waveguide`s orientation, while the solid curves are the respective fit-free simulations from COMSOL multiphysics. The color of each curve matches the color of the mechanical modes shown in \Cref{fig:Fig3}(a,b); \textbf{(f,g)} Brillouin gain and peak frequency as a function of the waveguide top width for the blue mode shown in \Cref{fig:Fig3}(d,e) for a fixed $\theta =$ \SI{0}{\degree}.}
    \label{fig:Fig3}
\end{figure*}

To unveil this polarization dependence, the BSBS response of our system is characterized using a pump-probe scheme, illustrated in \Cref{fig:Fig2}(c,d). A fixed-wavelength pump at $\lambda_{p}$ = \SI{1550}{\nm} and a piezo-scanned probe laser tuned around the pump frequency excites the acoustic waves. Inspired by previous experiments of SBS characterization~\cite{Grubbs1994,PhysRevLett.124.013902,doi:10.1126/sciadv.abq2196} we modulate the intensity of both the pump and probe lasers at slightly different frequencies ($f_{s}$ = $f_{p}$ + $\delta f$, where we set $f_{s}$ = \SI{3}{\MHz} and $\delta f$ = \SI{70}{\kHz}). This modulation allows us to detect the SBS gain signal at the difference frequency (\(\delta f\)) through a lock-in amplifier (LIA) system, which enhances the signal-to-noise ratio by mitigating other frequency components and minimizes the impact of the pump wave's spurious reflection from the probe signal (more details in \Cref{supp:Brillouin_Gain_Calibration} of Supplementary Material).

The polarization states of the pump and probe within the waveguide are adjusted and determined using an auxiliary optical ring resonator coupled to the waveguide. Although other methods exist for determining on-chip polarization, we chose the ring resonator because of its convenience and straightforward optical loss characterization. The microring transmission spectra, measured independently for the pump and probe waves, are presented in \Cref{fig:Fig2}e. To ensure unambiguous mode identification, we assess how the free spectral range varies with frequency~\cite{10.1063/5.0028839}, as illustrated in \Cref{fig:Fig2}f, which clearly distinguishes the polarization states of pump and probe waves by comparison with simulation results. Furthermore, the quality factors of the modes further confirm this identification, with the TE modes (\Cref{fig:Fig2}h) exhibiting sharper resonances compared to the TM modes (\Cref{fig:Fig2}g). More details can be found in \Cref{supp:Ring_Resonator_Optical_Modes} of Supplementary Material.

The spectrum of BSBS for each polarization combination is presented in \Cref{fig:Fig2}(i,j,k), providing clear evidence of strong cross-polarization scattering. A prominent peak around \SI{7.5}{\GHz} is observed when the pump and probe are launched in opposite polarization states (TE-TM or TM-TE), while a significantly weaker signal (around 10 dB weaker) is detected in the copolarized TE-TE configuration. The TM-TM configuration is not shown as no SBS signal was observed in this case. In the TE-TE configuration, BSBS exhibits a new peak at a slightly higher frequency of \SI{8}{\GHz}, although the \SI{7.5}{\GHz} signal, which dominates the cross-polarized spectra, is also measurable. We attribute this residual TE-TM peak in the TE-TE BSBS spectrum to finite-polarization extinction when pump and probe signals are launched using a lensed fiber. The average input powers at the chip facet for all configurations were 20 dBm for the pump and 10 dBm for the probe, while the coupling losses are estimated to be 7.2 dB per facet.

To understand the origins of strong cross-polarization BSBS and quantify the Brillouin gain, various waveguides with different widths and orientations on the same LNOI chip were analyzed. The BSBS gain spectra are shown in \Cref{fig:Fig3}(a,b) for two orientations relative to the LN's $Y$-axis, $\theta = $ \SI{0}{\degree} and $\theta = $ \SI{-20}{\degree}, respectively. 
As we deviate from $\theta = $ \SI{0}{\degree}, new modes begin to emerge (red curve), accompanied by a decrease in the frequency of the primary mode (blue curve), a consequence of the mechanical anisotropy. Since all waveguides have an inverted taper~\cite{Almeida:03} segment, which is aligned to the $Y$-axis, a residual peak at \SI{7.5}{\GHz} is present in all samples. This residual peak is indicated in green in \Cref{fig:Fig3}b. As the waveguide rotates around the $Z$-axis, the Brillouin spectrum evolves, as shown in \Cref{fig:Fig3}c. The resulting angular dispersion of the gain and frequency for each of these modes is illustrated in \Cref{fig:Fig3}(d,e), which showcases the well-known C$_{3}$ (threefold) symmetry inherent to the $Z$-cut LN crystal. Furthermore, there is excellent quantitative agreement between the experimental data and the predictions from a fit-free numerical simulation. The discrepancies in the absolute values could be optimized by fine-tuning specific material and geometric parameters, such as refractive index, stiffness, and piezoelectric coefficients. The blue mode exhibits a mechanical quality factor close to $Q_{m} \approx 100$, while the red mode has $Q_{m} \approx 20$. As we discuss later, these values are slightly lower than predicted by the acoustic radiation losses (around 270 and 75, respectively), suggesting that surface roughness, inhomogeneous broadening~\cite{Zurita:21,Wolff_2016}, and material absorption could play a role. This limited quality factor suggests that improved fabrication could increase the Brillouin gain $G_{B}$ in LNOI much above $\SI{80}{\m^{-1}\W^{-1}}$, a remarkably high value, only a factor of 4 lower than copolarized gains previously observed in chalcogenide glasses~\cite{Pant:11}.

The effect of optical and mechanical mode confinement on SBS gain was also analyzed at a fixed orientation of $\theta =$ \SI{0}{\degree} while varying the waveguide width, as shown in \Cref{fig:Fig3}(f,g). As the width of the waveguide increases, the gain decreases due to weaker mode overlap, while the frequency shifts to higher values due to increased effective stiffness $m_{\text{eff}}\Omega_{m}^{2}$, which is the product of effective mass $m_{\text{eff}}$ and the square of the mechanical frequency $\Omega_{m}$. This term appears in the denominator when analytically calculating the Brillouin gain, as discussed in \cite{10.1063/1.5088169}, which also explains that stiffer modes result in a lower gain. This phenomenon can be intuitively understood once the mechanical mode is distinctly identified, as highlighted in the subsequent discussion.

\begin{figure}[h!]
    \centering
    \includegraphics{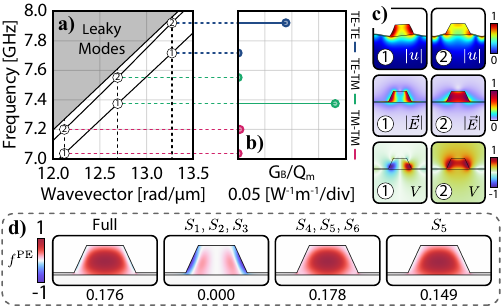}
    \caption{Finite element method (FEM) simulations for fixed $w = $ \SI{500}{\nm} and $\theta = $ \SI{0}{\degree} were performed using COMSOL multiphysics. \textbf{(a)} Mechanical dispersion; \textbf{(b)} Normalized Brillouin gains; \textbf{(c)} Piezoelectric modes profile, showing their acoustic displacement, RF electric field and electric potential field profiles. Quasistatic approximation was used; \textbf{(d)} Normalized photoelastic force densities ($f^{\text{PE}}$) comparison of mode labeled 1 under different conditions. (left to right) We first conducted the full simulation, then we compared with a scenario considering only the pressure components $S_{1}$, $S_{2}$, $S_{3}$ as nonzero and then only the shear components $S_{4}$, $S_{5}$, $S_{6}$ as nonzero to, finally, only consider the $S_{5}$ component. The bottom number shows the photoelastic normalized gain $G_{B}^{\text{PE}}/Q_{m}$ in $\SI{}{\m^{-1}\W^{-1}}$ for each case.}
    \label{fig:Fig4}
\end{figure}

To understand how such a strong cross-polarization BSBS can emerge in LNOI waveguides, the mechanical modes involved in our experiment must be identified. In \Cref{fig:Fig4}a we show the calculated mechanical dispersion relation of a $w = $ \SI{500}{\nm} waveguide aligned with the $Y$-axis. Only two mechanical modes lie outside the gray-shaded continuum of modes, where significant coupling to the silica substrate occurs, which is precisely the number of modes identified in our experimental data. Likewise, only the TE-TE and TE-TM polarizations exhibit a substantial Brillouin gain, as shown in \Cref{fig:Fig4}b. In the TM-TM case, no significant mode is excited, in agreement with the experiment. The combination TE-TM excites the fundamental SAW mode, producing the highest normalized Brillouin gain of $G_{B}/Q_{m} = $ \SI{0.183}{\m^{-1}\W^{-1}} around \SI{7.38}{\GHz}. The combination TE-TE appeared with a higher frequency around \SI{7.92}{\GHz}, which is consistent with our experimental findings in \Cref{fig:Fig2}k, and with a smaller normalized gain of $G_{B}/Q_{m} = $ \SI{0.084}{\m^{-1}\W^{-1}}. The difference frequency values between the TE-TM and TE-TE cases closely match the experimental separation of \SI{500}{\MHz} observed in \Cref{fig:Fig2}(i,j,k), which is an excellent agreement given the fit-free simulation parameters. The values of the material tensors are all shown in \Cref{supp:Material_Tensors_and_Piezoelectricity_Model} of Supplementary Material.

To couple orthogonally polarized optical modes, the acoustic mode must induce appreciable shear strain. For example, in optical fibers where interpolarization FSBS has been observed, the coupling occurs through the in-plane strain tensor component $S_{xy}$. The magnitude of the transverse shear stress is dominant in FSBS due to the small contribution of the longitudinal strain component $S_{zz} = \partial_{z} u_{z}\approx \mathrm{i} q_{m} u_{z}$, where $u_{z}$ is the displacement component along the propagation direction and $q_{m}$ is the acoustic wavenumber, which is close to zero in FSBS, that is, $q_{m} \approx 0$. For backward SBS, the phase-matching condition $q_{m} \approx 2 k_{p}$, where $k_{p}$ is the optical pump wavenumber, favors a strong longitudinal wave character that suppresses interpolarization scattering. This condition no longer holds for the SAW modes identified in \Cref{fig:Fig4}. The mode branch labeled 1 in \Cref{fig:Fig4}a is 58\% $\hat{x}$-polarized and 38\% $\hat{y}$-polarized, and is only excited in the cross-polarization configuration. In contrast, mode branch 2 is 84\% $\hat{y}$-polarized and is excited exclusively in the copolarized TE-TE configuration. Both modes are quasi-guided within the LN thin film, indicating that their velocities (3650 m/s and 3738 m/s) are very close to those of the silica substrate (3763 m/s). The guiding condition is slightly frustrated by the radiation of the SAWs, as a thin residual LN film remains present, leading to radiation-limited mechanical quality factors.

In \Cref{fig:Fig4}d we show the density of the photoelastic optical force~\cite{10.1063/1.5088169}, which dominates the SBS gain, compared to moving boundary and roto-optical~\cite{doi:10.1121/1.390547,eggleton2022brillouin} interactions. The contribution of the longitudinal strain~\cite{auld1973acoustic} components $S_{1}, S_{2}$ and $S_{3}$, which are usually responsible for standard intramodal SBS interactions, is zero. The major contribution to the photoelastic force density is due to the shear strain components $S_{4}, S_{5}$ and $S_{6}$, but specially $S_{5}=(S_{xz}+S_{zx})/2$, which account for 85\% of the SBS gain. The contribution of this term to the force density is given by
\begin{equation}
     f^\text{PE} \propto \left[\left(D^{p}_{x}\right)^{*}D^{s}_{y} + \left(D^{p}_{y}\right)^{*}D^{s}_{x}\right]p_{41}S_{5},
\end{equation}
where $D^{(p,s)}_{x,y}$ are the pump ($p$) and Stokes ($s$) Cartesian electric displacement field components of the optical modes and $p_{41}=-0.151$ is the photoelastic tensor component.
This is rather unique to LNOI, as $p_{41}$ is zero in all isotropic materials. Thus, not only does this platform guide shear SAW modes but also enhances their participation in BSBS mainly through the large nontrivial $p_{41}$ component, present only in anisotropic materials. Finally, the shear nature of the mechanical mode highlighted in \Cref{fig:Fig4}d provides an intuitive explanation for the frequency increase observed in \Cref{fig:Fig3}g. The torsional stiffness of a rectangular cross-section bar increases linearly with its width, which qualitatively accounts for the observed frequency increase.

It should be emphasized that piezoelectric coupling in LN was considered for all calculations in mechanical mode, unlike our previous study~\cite{Rodrigues:23} which ignored it. Without this coupling, qualitative insight into Brillouin optomechanics in LNOI is possible but lacks quantitative precision. The high piezoelectric coupling in LN leads to coupled electromechanical modes, as shown in \Cref{fig:Fig4}c. Although our focus is not on the RF interface~\cite{Liu:19,wang2024noncontactexcitationmultighzlithium,snigirev2022ultrafasttunablelasersusing,Xie2023}, the mechanical modes could be used in future experiments with short-period IDTs for microwave-to-optical transduction through Brillouin scattering~\cite{Zhou2024,PhysRevResearch.5.043140}, given the spatial profile of the electric potential shown in \Cref{fig:Fig4}c. Performing such an interface on an anchored waveguide could improve the thermalization of the system, affecting quantum regime operations~\cite{Jiang:19,mayor2024twodimensionaloptomechanicalcrystalquantum}.

In conclusion, this study demonstrates cross-polarized backward stimulated Brillouin scattering in lithium niobate on insulator waveguides, highlighting significant SBS gains. This opens up pathways for high-performance photonic components like ultranarrowband lasers, broadband nonreciprocal devices, and RF photonic filters that may leverage the polarization degree of freedom. For instance, in SBS laser applications~\cite{Gundavarapu2019} or Brillouin-based quantum-state preparation~\cite{Vanner2015TowardsMotion}, a polarizer could be used to enhance the suppression of undesired pump photons. The observed Brillouin gain in the cross-polarization regime showcases LNOI's unique capabilities, suggesting integration with RF piezoelectric actuation for novel polarization-based photonic devices. These findings indicate the potential for SBS in LNOI to transform polarization-sensitive optical processing, advancing integrated photonics and optomechanics for scalable and efficient applications.
\\

\noindent \textbf{Note}. During the preparation of this manuscript, we became aware of the work by Kaixuan Ye et al.~\cite{ye2023surface} also reporting SBS in LNOI.\\

\noindent \textbf{Funding}. This work was supported by the São Paulo Research Foundation (FAPESP) through Grants
19/14377-5, 
22/11486-0, 
22/06254-3, 
19/13564-6, 
23/01206-3, 
18/15577-5, 
18/15580-6, 
18/25339-4, 
Coordenação de Aperfeiçoamento de Pessoal de Nível Superior - Brasil (CAPES) (Financial Code 001), Conselho Nacional de Desenvolvimento Científico e Tecnológico (CNPQ) Grant 409626/2022-8; and Fulbright-CAPES Grant 88881.625368/2021-01. Further support was provided by the Defense Advanced Research Projects Agency (DARPA) (HR0011-20-C-0137) and the Harvard University Dean’s Competitive Fund for Promising Scholarship.\\

\noindent \textbf{Disclosures}.
The authors declare no conflict of interest.\\

\noindent \textbf{Data availability}. The data and simulation files reported in the manuscript will be available upon publication in ZENODO (10.5281/zenodo.10059309).

\bibliography{references_new.bib}

\newpage

\begin{widetext}

\renewcommand{\arraystretch}{1.25}

\newpage

\section*{Supplementary Material}

\subsection{Sample Details and Fabrication Recipe}\label{supp:Sample_Details_and_Fabrication_Recipe}

The fabrication process begins with a clean piece of LNOI (\SI{400}{\nm} thick LN) subjected to a 1-minute sonication in acetone, followed by rinsing with IPA and DI water, and then a 5-minute acid piranha treatment (3 parts H$_{2}$SO$_{4}$ : 1 part H$_{2}$O$_{2}$ at self-heating) to ensure the elimination of any organic residues and metal contaminants. After the acid piranha treatment, we apply an adhesion promoter (SurPass 3000) since ma-N resist does not adhere well to LN. We then coat the sample using ma-N 2405 with the spinner at 3000 rpm for 45 seconds, achieving a thickness of approximately \SI{500}{\nm}, subsequently baking it for 1 minute at \SI{90}{\celsius}. The patterning is done using electron beam lithography with \SI{50}{\kV} acceleration voltage (Elionix HS50). The total dose was 360 $\mu$C/cm$^2$ at a constant current of \SI{1}{\nA}, utilizing multipass exposure techniques (four passes total) combined with overlapping writing fields to minimize both roughness and stitching errors. Employing Ar$^+$ plasma in the ICP-RIE tool (ULVAC NLD-570), we dry etch the LN thin film until we almost reach the silica layer. Subsequently, to prevent redeposition and other contaminants by post-etching, the sample undergoes a 10-minute immersion in SC1 solution (5 parts DI water: 1 part NH$_{4}$OH: 1 part H$_{2}$O$_{2}$) at \SI{60}{\celsius}, followed again by a 5-minute treatment in acid piranha at self-heating temperatures. We conclude the fabrication process by annealing the sample in an O$_{2}$ atmosphere at \SI{520}{\celsius} overnight~\cite{10.1063/5.0095146}.

The sample design was influenced by our prior research~\cite{Rodrigues:23}. Due to the scarcity of experimental articles on SBS in LNOI~\cite{ye2023surface,s11433-023-2272-y}, our design aimed to measure the angular dependence of SBS in relation to the crystal's anisotropy and explore intermodal scattering. This led to the distinctive ``flower shape" design of the sample, as depicted in \Cref{fig:Fig1_Supp}a. Photo of our experimental setup, highlighting coupling through lensed fibers, with the $X$, $Y$, and $Z$ stage controls are shown in \Cref{fig:Fig1_Supp}(b,c).

\begin{figure}[ht]
    \centering
    \includegraphics{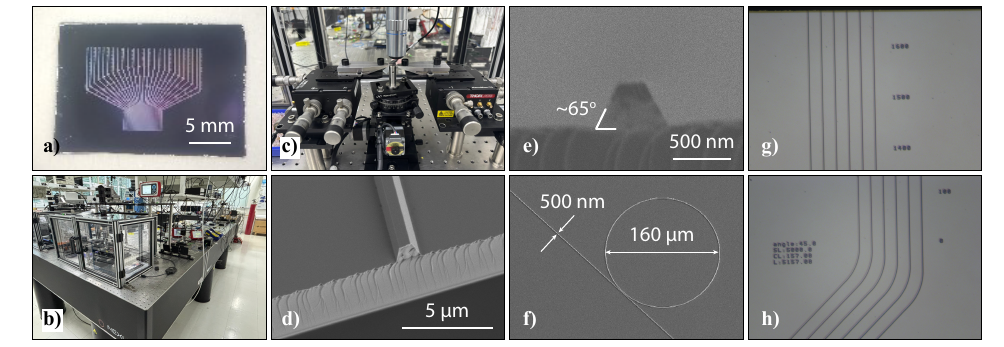}
    \caption{\textbf{(a)} Image of the fabricated sample before cleavage (GDS file available on Zenodo); \textbf{(b,c)} Photos of our laboratory, showing part the setup used for the article; \textbf{(d,e,f)} Scanning electron microscope (SEM) images showcasing various perspectives of the sample, such as the cross-section, the ring resonator, and the cleaved segment which connects with the lensed fiber; \textbf{(g,h)} Optical images of the sample providing insight into the measurement of the waveguide length using a ruler as well as the label in the corner section which is used to identify the waveguide.}
    \label{fig:Fig1_Supp}
\end{figure}

The waveguide families range from $\theta = \SI{-60}{\degree}$ to $\theta = \SI{60}{\degree}$, in steps of \SI{5}{\degree} relative to the LN $Y$-axis, with $\theta = 0$ indicating propagation along the $Y$-axis. Each orientation is accompanied by waveguide top widths ranging from $w =$ \SI{500}{\nm} to $w =$ \SI{1000}{\nm} in steps of \SI{100}{\nm}. These waveguides are standardized with a length of \SI{5}{\mm} (excluding the inverted taper segment) and have a wedge angle of $\phi=$ \SI{65}{\degree} (see \Cref{fig:Fig1_Supp}e) with a remaining slab thickness of $\delta t=$ \SI{30}{\nm}. To ensure accurate length calibration post-cleaving, a ruler is placed on the side of each waveguide, as shown in \Cref{fig:Fig1_Supp}(g,h). Each waveguide family is labeled (seen on the left side of \Cref{fig:Fig1_Supp}h) to facilitate tracking. The exact lengths for each waveguide family are detailed in \Cref{tab:L2_values}.

The LNOI wafer thickness varies by $\pm$\SI{10}{\nm} along the propagation length of our devices, which is not taken into account in our simulations or fabrication process and may affect the comparisons of experiments and theory. We used an average thickness $t=$ \SI{400}{\nm} and a slab thickness $\delta t=$ \SI{30}{\nm}, both estimated from Filmetrics and SEM measurements. The wedge angle $\phi = $ \SI{65}{\degree} was measured by SEM (\Cref{fig:Fig1_Supp}e) and validated with COMSOL simulations that match the group index found in the microring resonator. The refractive index and the absorption properties of LN were measured by ellipsometry and their values are shown in \Cref{fig:Fig2_Supp}.

\begin{figure}[ht]
    \centering
    \includegraphics{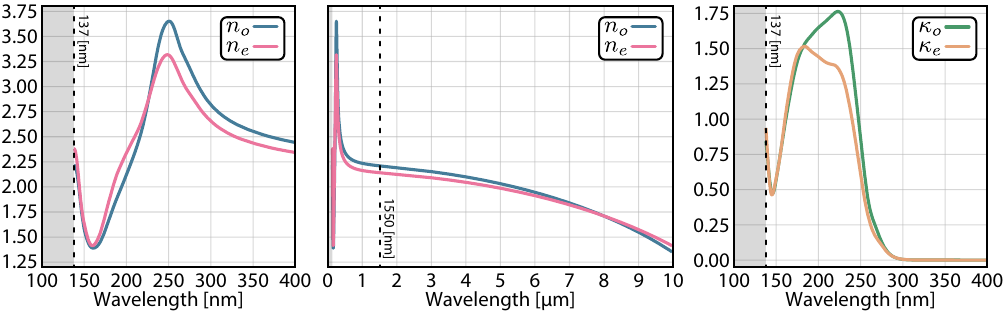}
    \caption{Complex refractive indices of lithium niobate, which covered a wavelength range from \SI{137}{\nm} to \SI{10}{\um}. The measurements of $\kappa_{o}$ and $\kappa_{e}$ were zero for wavelengths exceeding \SI{400}{\nm}.}
    \label{fig:Fig2_Supp}
\end{figure}

We fabricated the system with a remaining slab because, as reported in~\cite{Rodrigues:23}, it is not critical for phonon leakage and allows the integration of electrodes onto the LN thin film, as done for LN electro-optical devices~\cite{Zhang:21,Yu2023}. Reporting a large Brillouin gain on LNOI devices with a remaining slab is thus appealing. The cross-section profile of the inverse taper is shown in \Cref{fig:Fig1_Supp}(c,e). Some waveguides also feature a ring resonator with a radius of \SI{80}{\um}, as depicted in \Cref{fig:Fig1_Supp}f, essential for identifying the optical polarization of the pump and probe used in the experiment.

\subsection{Material Tensors and Piezoelectricity Model}\label{supp:Material_Tensors_and_Piezoelectricity_Model}

We used the Stress-Charge formulation of piezoelectricity, as defined by

\begin{equation}\label{equacao_piezoeletricidade}
    \begin{cases}
        \mathbf{T} = \mathbf{C}^{\textbf{E}}:\mathbf{S} - \mathbf{e}^{T}\cdot\mathbf{E} \\
        \mathbf{D} = \mathbf{\varepsilon}^{\textbf{S}}\cdot\mathbf{E} + \mathbf{e}:\mathbf{S}
    \end{cases}
    \quad \Rightarrow \quad
    \begin{cases}
        T_{I} = C_{IJ}^{\textbf{E}}S_{J} - e_{Ij}E_{j} \\
        D_{i} = \varepsilon_{ij}^{\textbf{S}}E_{j} + e_{iJ}S_{J}
    \end{cases}
\end{equation}

where $\mathbf{e}$ denotes the piezoelectric tensor, $\mathbf{T}$ is the stress tensor, $\mathbf{S}$ is the strain tensor, $\mathbf{E}$ represents the electric field and $\mathbf{D}$ is the electric displacement field. The stiffness tensor $\mathbf{C}^{\textbf{E}}$ and the electric permittivity tensor $\mathbf{\varepsilon}^{\textbf{S}}$ are given under conditions of constant electric field and constant strain field, respectively. In \Cref{equacao_piezoeletricidade}, uppercase indices (1 to 6) represent Voigt notation, while lowercase indices (1 to 3) refer to Cartesian components. The relevant material tensors are shown below: $C_{IJ}^{\textbf{E}}$ is in units of [GPa], $e_{iJ}$ is in units of [C/m$^{2}$], the electric permittivity is normalized by the vacuum permittivity $\varepsilon_{0}$, making $\varepsilon_{ij}^{\textbf{S}}/\varepsilon_{0}$ dimensionless (do not confuse the RF permittivity with the optical electric permittivity values presented in \Cref{fig:Fig2_Supp}). We also include the photoelastic tensor $p_{IJ}$. All tensors are represented on the crystallographic basis~\cite{ma15134716,Weis1985,cryst11091095,LEDBETTER2004941,Yamada_1967,doi:10.1063/1.1660528}.

\begin{equation}
    \left[\varepsilon_{ij}^{\textbf{S}}/\varepsilon_{0}\right] =
    \begin{pmatrix}
        \varepsilon_{o}^{\textbf{S}}/\varepsilon_{0} & 0 & 0\\
        0 & \varepsilon_{o}^{\textbf{S}}/\varepsilon_{0} & 0\\
        0 & 0 & \varepsilon_{e}^{\textbf{S}}/\varepsilon_{0}
    \end{pmatrix}
    =
    \begin{pmatrix}
        44.95 & 0 & 0\\
        0 & 44.95 & 0\\
        0 & 0 & 24.06
    \end{pmatrix}
\end{equation}

\begin{equation}
    \left[C_{IJ}^{\textbf{E}}\right] =
    \begin{pmatrix}
        c_{11}^{\textbf{E}} & c_{12}^{\textbf{E}} & c_{13}^{\textbf{E}} & c_{14}^{\textbf{E}} & 0 & 0\\
        c_{12}^{\textbf{E}} & c_{11}^{\textbf{E}} & c_{13}^{\textbf{E}} & -c_{14}^{\textbf{E}} & 0 & 0\\
        c_{13}^{\textbf{E}} & c_{13}^{\textbf{E}} & c_{33}^{\textbf{E}} & 0 & 0 & 0\\
        c_{14}^{\textbf{E}} & -c_{14}^{\textbf{E}} & 0 & c_{44}^{\textbf{E}} & 0 & 0\\
        0 & 0 & 0 & 0 & c_{44}^{\textbf{E}} & c_{14}^{\textbf{E}}\\
        0 & 0 & 0 & 0 & c_{14}^{\textbf{E}} & \dfrac{c_{11}^{\textbf{E}}-c_{12}^{\textbf{E}}}{2}
    \end{pmatrix}
    =
    \begin{pmatrix}
        198.83 & 54.64 & 68.23 & 7.83 & 0 & 0\\
        54.64 & 198.83 & 68.23 & -7.83 & 0 & 0\\
        68.23 & 68.23 & 235.71 & 0 & 0 & 0\\
        7.83 & -7.83 & 0 & 59.86 & 0 & 0\\
        0 & 0 & 0 & 0 & 59.86 & 7.83\\
        0 & 0 & 0 & 0 & 7.83 & 72.095
    \end{pmatrix}
\end{equation}

\begin{equation}
    \left[e_{iJ}\right] =
    \begin{pmatrix}
        0 & 0 & 0 & 0 & e_{15} & -e_{22} &\\
        -e_{22} & e_{22} & 0 & e_{15} & 0 & 0 &\\
        e_{31} & e_{31} & e_{33} & 0 & 0 & 0
    \end{pmatrix}
    =
    \begin{pmatrix}
        0 & 0 & 0 & 0 & 3.65 & -2.39 &\\
        -2.39 & 2.39 & 0 & 3.65 & 0 & 0 &\\
        0.31 & 0.31 & 1.72 & 0 & 0 & 0
    \end{pmatrix}
\end{equation}

\begin{equation}
    \left[p_{IJ}\right] =
    \begin{pmatrix}
        p_{11} & p_{12} & p_{13} & p_{14} & 0 & 0\\
        p_{12} & p_{11} & p_{13} & -p_{14} & 0 & 0\\
        p_{31} & p_{31} & p_{33} & 0 & 0 & 0\\
        p_{41} & -p_{41} & 0 & p_{44} & 0 & 0\\
        0 & 0 & 0 & 0 & p_{44} & p_{41}\\
        0 & 0 & 0 & 0 & p_{14} & \dfrac{p_{11}-p_{12}}{2}
    \end{pmatrix}
    =
    \begin{pmatrix}
        -0.026 & 0.088 & 0.134 & -0.083 & 0 & 0\\
        0.088 & -0.026 & 0.134 & 0.083 & 0 & 0\\
        0.177 & 0.177 & 0.07 & 0 & 0 & 0\\
        -0.151 & 0.151 & 0 & 0.145 & 0 & 0\\
        0 & 0 & 0 & 0 & 0.145 & -0.151\\
        0 & 0 & 0 & 0 & -0.083 & -0.057
    \end{pmatrix}
\end{equation}

\subsection{Brillouin Gain Calibration}\label{supp:Brillouin_Gain_Calibration}

To calculate the LN Brillouin gain ($G_{B}^{\text{LN}}$), we used the Brillouin gain of the silica pigtail as a reference~\cite{doi:10.1126/sciadv.abq2196}. This decision was due to the inevitability of silica fiber within our system, which connects all devices. The part of the setup relevant for the calibration is shown in \Cref{fig:Fig3_Supp}a, and the key parameters for the calculations are defined in \Cref{fig:Fig3_Supp}b. 

\begin{figure*}[ht]
    \centering
    \includegraphics{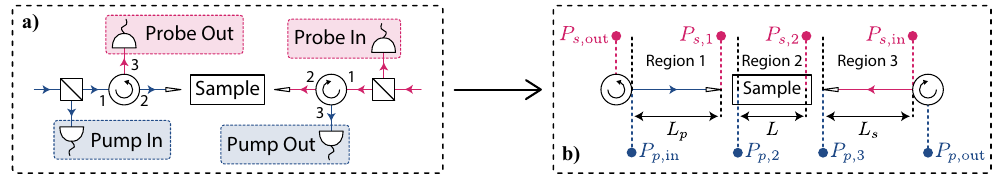}
    \caption{\textbf{(a)} A segment of the experimental setup, highlighting the region adjacent to the sample crucial for Brillouin gain calibration; \textbf{(b)} Key parameters essential for the algebraic calculation of Brillouin gain calibration.}
    \label{fig:Fig3_Supp}
\end{figure*}

Assuming the pump (subscript $p$) is not depleted by the Stokes signal (subscript $s$), the equations for the pump power are

\begin{align}\label{Pump_powers}
    \begin{cases}
        P_{p,2}(t) &= T_{p}^{-}P_{p,\text{in}}(t)\\
        P_{p,3}(t) &= T_{p}^{+}P_{p,2}(t)e^{-\alpha_{p}L}\\
        P_{p,\text{out}}(t) &= P_{p,3}(t)
    \end{cases}
    \quad \Rightarrow \quad
    \begin{cases}
        P_{p,2}(t) &= T_{p}^{-}P_{p,\text{in}}(t)\\
        P_{p,3}(t) &= T_{p}^{+}T_{p}^{-}P_{p,\text{in}}(t)e^{-\alpha_{p}L}\\
        P_{p,\text{out}}(t) &= T_{p}^{+}T_{p}^{-}P_{p,\text{in}}(t)e^{-\alpha_{p}L}
    \end{cases}
\end{align}

We ignore fiber propagation losses (about 0.2 dB/km at \SI{1550}{\nm}) due to the use of only 15 meters of silica fiber patch cords. The parameters $T_{p}^{+}$ and $T_{p}^{-}$ denote the pump transmittance on the right and left facets, respectively. The parameter $\alpha_{p}$ represents the loss of propagation of the pump mode within the sample, and $L$ is the length of the waveguide. The Stokes signal equations share a similar structure and can be solved as a function of $P_{p,\text{in}}(t)$ and $P_{s,\text{in}}(t)$ as

\begin{multline}\label{Probe_powers}
    \begin{cases}
        P_{s,2}(t) &= T_{s}^{+}P_{s,\text{in}}(t)e^{G_{B,3}P_{p,3}(t)L_{s}}\\
        P_{s,1}(t) &= T_{s}^{-}P_{s,2}(t)e^{G_{B,2}P_{p,2}(t)L}e^{-\alpha_{s}L}\\
        P_{s,\text{out}}(t) &= P_{s,1}(t)e^{G_{B,1}P_{p,\text{in}}(t)L_{p}}
    \end{cases}
    \quad \Rightarrow\\
    \Rightarrow \quad
    \begin{cases}
        P_{s,2}(t) &= T_{s}^{+}P_{s,\text{in}}(t)e^{G_{B,3}T_{p}^{+}T_{p}^{-}P_{p,\text{in}}(t)L_{s}e^{-\alpha_{p}L}}\\
        P_{s,1}(t) &= T_{s}^{-}T_{s}^{+}P_{s,\text{in}}(t)e^{G_{B,3}T_{p}^{+}T_{p}^{-}P_{p,\text{in}}(t)L_{s}e^{-\alpha_{p}L}}e^{G_{B,2}T_{p}^{-}P_{p,\text{in}}(t)L}e^{-\alpha_{s}L}\\
        P_{s,\text{out}}(t) &= T_{s}^{-}T_{s}^{+}P_{s,\text{in}}(t)\underbrace{e^{G_{B,3}T_{p}^{+}T_{p}^{-}P_{p,\text{in}}(t)L_{s}e^{-\alpha_{p}L}}}_\text{Region 3}\underbrace{e^{G_{B,2}T_{p}^{-}P_{p,\text{in}}(t)L}e^{-\alpha_{s}L}}_{\text{Region 2}}\underbrace{e^{G_{B,1}P_{p,\text{in}}(t)L_{p}}}_{\text{Region 1}}
    \end{cases}
\end{multline}

Similarly, $T_{s}^{+}$ and $T_{s}^{-}$ denote the transmittance of the Stokes signal in the right and left facets respectively; $\alpha_{s}$ is the propagation loss of the Stokes mode; $L_{p}$ is the length that the pump input travels from one of the circulator to the lensed fiber, while $L_{s}$ is the length that the probe signal input travels from the other circulator to the other lensed fiber. The Brillouin gain coefficients $G_{B,1}$, $G_{B,2}$, and $G_{B,3}$ are functions of the frequency difference between the pump and probe signals. If the frequency difference of pump and probe is such that it matches the silica fiber Brillouin mode, then $G_{B,1} = G_{B,3} = G_{B}^{\text{fiber}}$ and $G_{B,2} = 0$, as we do not have a silica fiber Brillouin mode inside the LNOI chip. Thus, the output power for this case, which we will call $P_{s,\text{out}}^{\text{fiber}}(t)$, is given by

\begin{equation}\label{Pout_silica}
    P_{s,\text{out}}^{\text{fiber}}(t) = T_{s}^{-}T_{s}^{+}P_{s,\text{in}}(t)e^{G_{B}^{\text{fiber}}P_{p,\text{in}}(t)\left(L_{p}+T_{p}^{+}T_{p}^{-}L_{s}e^{-\alpha_{p}L}\right)}e^{-\alpha_{s}L}
\end{equation}

If the frequency difference is now such that it excites the on-chip LN waveguide Brillouin mode, then $G_{B,1} = G_{B,3} = 0$ and $G_{B,2} = G_{B}^{\text{LN}}$, and such output power $P_{s,\text{out}}^{\text{LN}}(t)$ is given by

\begin{equation}\label{Pout_LN}
    P_{s,\text{out}}^{\text{LN}}(t) = T_{s}^{-}T_{s}^{+}P_{s,\text{in}}(t)e^{G_{B}^{\text{LN}}T_{p}^{-}P_{p,\text{in}}(t)L}e^{-\alpha_{s}L}
\end{equation}

The optical powers $P_{p,\text{in}}(t)$ and $P_{s,\text{in}}(t)$ were intensity modulated at the quadrature points, a method previously used to measure small SBS gains in silicon nitride~\cite{doi:10.1126/sciadv.abq2196,PhysRevLett.124.013902,botter2023stimulated}. Thus, $P_{p,\text{in}}(t)$ and $P_{s,\text{in}}(t)$ can be written~\cite{10.5555/1817101} as a Jacobi-Anger expansion

\begin{align}\label{quadrature_point_p}
    P_{p,\text{in}}(t) &= P_{p,\text{in}}^{\text{DC}}\left(1 + 2\sum_{n=1}^{\infty}J_{2n-1}\left(\varepsilon_{p}\right)\sin\left[\left(2n-1\right)\left(\Omega_{p} t + \theta_{p}\right)\right]\right), \quad \text{where} \quad \varepsilon_{p} := \pi\frac{V_{\text{AC}}^{(p)}}{V_{\pi}^{(p)}}
\end{align}

\begin{align}\label{quadrature_point_s}
    P_{s,\text{in}}(t) &= P_{s,\text{in}}^{\text{DC}}\left(1 + 2\sum_{n=1}^{\infty}J_{2n-1}\left(\varepsilon_{s}\right)\sin\left[\left(2n-1\right)\left(\Omega_{s} t + \theta_{s}\right)\right]\right), \quad \text{where} \quad \varepsilon_{s} := \pi\frac{V_{\text{AC}}^{(s)}}{V_{\pi}^{(s)}}
\end{align}

The terms $P_{p,\text{in}}^{\text{DC}}$ and $P_{s,\text{in}}^{\text{DC}}$ denote the pump and probe DC optical power; $J_{n}(\varepsilon_{p})$ and $J_{n}(\varepsilon_{s})$ are the $n$-th order Bessel functions of the first kind for modulation depths $\varepsilon_{p}$ and $\varepsilon_{s}$. Modulations have angular frequencies given by $\Omega_{p}$ and $\Omega_{s}$ and phases $\theta_{p}$ and $\theta_{s}$. Modulation of both signals is essential for reasons in~\cite{PhysRevLett.124.013902}:

\begin{itemize}
    \item These modulations enable experiments outside the noisy ``DC World'';
    \item They filter out the pump reflection from the probe signal;
\end{itemize}

We acknowledge the impact of Kerr and Raman effects, but do not use a third laser as in~\cite{PhysRevLett.124.013902} to cancel them out. Because of the higher Brillouin gain in LN compared to that in silicon nitride, we believe that addressing these issues is unnecessary. We can approximate \Cref{quadrature_point_p} and \Cref{quadrature_point_s} to the first order as follows

\begin{equation}\label{Pp_Ps_mod_first_order}
    P_{p,\text{in}}(t) \approx P_{p,\text{in}}^{\text{DC}}\left[1 + 2J_{1}\left(\varepsilon_{p}\right)\sin\left(\Omega_{p}t + \theta_{p}\right)\right] \quad \text{and} \quad P_{s,\text{in}}(t) \approx P_{s,\text{in}}^{\text{DC}}\left[1 + 2J_{1}\left(\varepsilon_{s}\right)\sin\left(\Omega_{s}t + \theta_{s}\right)\right]
\end{equation}

This approximation holds if we employ small modulation depths. In our scenario, we have $\varepsilon_{p} \approx 1.05$ and $\varepsilon_{s} \approx 0.8$, ensuring that the ratio between consecutive terms in the expansion is less than 5\%, i.e.,

\begin{equation}
    \frac{J_{3}(\varepsilon_{p})}{J_{1}(\varepsilon_{p})} \approx 0.049 \quad \text{and} \quad \frac{J_{3}(\varepsilon_{s})}{J_{1}(\varepsilon_{s})} \approx 0.027
\end{equation}

As we demodulate at the frequency difference $\Delta\Omega = \Omega_{p} - \Omega_{s}$, it is crucial to isolate this particular spectral component from both \Cref{Pout_silica} and \Cref{Pout_LN}. To achieve this, we begin by expanding the exponential in \Cref{Pout_silica} and \Cref{Pout_LN} to the first order and then substitute \Cref{Pp_Ps_mod_first_order}. The resulting spectral component of interest (i.e., the components oscillating at the difference $\Omega_{p}-\Omega_{s}$) are expressed as

\begin{equation}
    P_{s,\text{out}}^{\text{fiber}}(t) = 2T_{s}^{-}T_{s}^{+}P_{p,\text{in}}^{\text{DC}}P_{s,\text{in}}^{\text{DC}}e^{-\alpha_{s}L}J_{1}(\varepsilon_{p})J_{1}(\varepsilon_{s})G_{B}^{\text{fiber}}\left(L_{p}+T_{p}^{+}T_{p}^{-}L_{s}e^{-\alpha_{p}L}\right)\cos{\left[\left(\Omega_{p}-\Omega_{s}\right)t + \left(\theta_{p}-\theta_{s}\right)\right]}
\end{equation}

\begin{equation}
    P_{s,\text{out}}^{\text{LN}}(t) = 2T_{s}^{-}T_{s}^{+}P_{p,\text{in}}^{\text{DC}}P_{s,\text{in}}^{\text{DC}}e^{-\alpha_{s}L}J_{1}(\varepsilon_{p})J_{1}(\varepsilon_{s})G_{B}^{\text{LN}}T_{p}^{-}L\cos{\left[\left(\Omega_{p}-\Omega_{s}\right)t + \left(\theta_{p}-\theta_{s}\right)\right]}
\end{equation}

The output signal of the photoreceiver (New Focus Model:2053-FS-M) corresponds to these same optical powers multiplied by a conversion factor $H_{s}$. The root mean square (RMS) of the photocurrent $I_{\text{RMS}}$ measured by our PXA signal analyzer is then determined as

\begin{align}
    &I_{\text{RMS}}^{\text{fiber}} := \text{RMS}\left[H_{s}^{\text{fiber}}P_{s,\text{out}}^{\text{fiber}}(t)\right] = \frac{H_{s}^{\text{fiber}}}{\sqrt{2}}2T_{s}^{-}T_{s}^{+}P_{p,\text{in}}^{\text{DC}}P_{s,\text{in}}^{\text{DC}}e^{-\alpha_{s}L}J_{1}(\varepsilon_{p})J_{1}(\varepsilon_{s})G_{B}^{\text{fiber}}\left(L_{p}+T_{p}^{+}T_{p}^{-}L_{s}e^{-\alpha_{p}L}\right)\\
    &I_{\text{RMS}}^{\text{LN}} := \text{RMS}\left[H_{s}^{\text{LN}}P_{s,\text{out}}^{\text{LN}}(t)\right] = \frac{H_{s}^{\text{LN}}}{\sqrt{2}}2T_{s}^{-}T_{s}^{+}P_{p,\text{in}}^{\text{DC}}P_{s,\text{in}}^{\text{DC}}e^{-\alpha_{s}L}J_{1}(\varepsilon_{p})J_{1}(\varepsilon_{s})G_{B}^{\text{LN}}T_{p}^{-}L
\end{align}

The ratio of these two quantities is

\begin{equation}\label{G_B_LN}
    \frac{I_{\text{RMS}}^{\text{LN}}}{I_{\text{RMS}}^{\text{fiber}}} = \frac{H_{s}^{\text{LN}}}{H_{s}^{\text{fiber}}}\frac{G_{B}^{\text{LN}}}{G_{B}^{\text{fiber}}}\frac{T_{p}^{-}L}{\left(L_{p}+T_{p}^{+}T_{p}^{-}L_{s}e^{-\alpha_{p}L}\right)} \quad \Rightarrow \quad G_{B}^{\text{LN}} \approx G_{B}^{\text{fiber}}\frac{\left(L_{p}+T_{p}^{+}T_{p}^{-}L_{s}e^{-\alpha_{p}L}\right)}{T_{p}^{-}L}\frac{I_{\text{RMS}}^{\text{LN}}}{I_{\text{RMS}}^{\text{fiber}}}
\end{equation}

In which the last approximation consists of assuming a flat gain for the optical detector across the frequency range, i.e. $H_{s}^{\text{fiber}} \approx H_{s}^{\text{LN}}$, thus eliminating the need for individual detector gain adjustments. The calculation of $G_{B}^{\text{LN}}$ does not depend on the calibration of the intensity modulator (modulation depth $\varepsilon_{p}$ and $\varepsilon_{s}$) or the DC level of each optical signal (pump and probe). Instead, it relies on factors such as silica fiber gain, sample losses, lengths, and the signal ratio. This method reduces experimental complexity and potential errors, similar to calculating $g_{0}$ in optomechanical cavities~\cite{Gorodetksy:10}, and has been explored previously~\cite{doi:10.1126/sciadv.abq2196}. The lengths of the fiber segments $L_{p}$ and $L_{s}$ are measured using an OTDR (Optical Time-Domain Reflectometer), while $L$ is predetermined by design and measured using the Fabry-Perot background of the transmitted signal. The propagation loss of the pump, $\alpha_{p}$, is calculated using the optical modes of the microring resonator. Assuming $T_{p}^{+} \approx T_{p}^{-} := T_{p}$, the pump transmittance is determined using the readings of the optical power meter and $\alpha_{p}$. Finally, $G_{B}^{\text{fiber}}$ is known from standard SMF-28 fibers but was calibrated for our specific setup.

\subsection{Gain Sensitivity to Transmittance Asymmetry}

We were concerned about the potential for highly asymmetric transmittances from the facets of the waveguides, which could introduce significant errors in the Brillouin gain calculation. This asymmetry may be caused not only by actual coupling losses but also from accidental scattering spots present in the uncladded waveguides (e.g., lithography faults or dust particles), or even misalignment during sample cleaving. A preliminary estimate of this gain sensitivity can be modeled using $T_{p}^{+}$ and $T_{p}^{-}$ as

\begin{equation}
    T_{p}^{+} = \rho T_{p} \quad \text{and} \quad T_{p}^{-} = T_{p}/\rho
\end{equation}

where $\rho \in (0,\infty)$ is the asymmetry parameter, in which $\rho=1$ means perfect symmetric. We can write $G_{B}^{\text{LN}}(\rho)$ as a function of $\rho$ and $G_{B}^{\text{LN}}(1)$ using \Cref{G_B_LN} as

\begin{equation}
    G_{B}^{\text{LN}}(\rho) = \rho G_{B}^{\text{LN}}(1) \quad \text{where} \quad G_{B}^{\text{LN}}(1) = G_{B}^{\text{fiber}}\frac{\left(L_{p}+T_{p}^{2}L_{s}e^{-\alpha_{p}L}\right)}{T_{p}L}\frac{I_{\text{RMS}}^{\text{LN}}}{I_{\text{RMS}}^{\text{fiber}}}.
\end{equation}

We observe that $G_{B}^{\text{LN}}(\rho)$ varies linearly with $\rho$ in our model, which is not a small dependency. As an example, let us suppose that one of the facets has twice the transmittance of the other facet, that is, we have $T_{p}^{+} = 2T_{p}^{-}$, which implies that $\rho = \sqrt{2}$. This value of $\rho$ means that $G_{B}^{\text{LN}}(\rho = \sqrt{2}) = \sqrt{2}G_{B}^{\text{LN}}(1) \approx 1.4 G_{B}^{\text{LN}}(1)$, which is about 40$\%$ of error on the amplitude of the Brillouin gain calculated, which could perfectly explain the behavior observed in \Cref{fig:Fig3}d at $\theta =$ \SI{-40}{\degree} where we expected a smaller gain value, making the curve more symmetric, consistent with the known C$_{3}$ symmetry of the LN crystal. Fortunately, this is not the case for almost all of our waveguides, as the facets are almost identical, and then to assume $\rho = 1$ is reasonable. Any imprecision on the facet should, necessarily, be on the order of the cross-section area shown in \Cref{fig:Fig1_Supp}d, which is on the order of few hundreds of nanometers at maximum, which is practically invisible for the optical wavelength used in the experiment of $\lambda_{p} = $ \SI{1550}{\nm}.

Another factor that would help in future experiments is to ensure that $G_{B}^{\text{LN}}(1)$ is independent of $T_{p}^{-}$, removing an extra imprecision on $\alpha_{p}$ and making the Brillouin gain a function of only one facet transmittance. This can be achieved forcing \(L_{p}/L_{s} \rightarrow \infty\), which means that the input pump fiber should be much longer than the input probe fiber, enhancing reliability while comparing waveguides with different cleaved facets and decreasing the overall error in the gain calculation. Moreover, the condition is actually given by $L_{p} \gg T_{p}^{2}L_{s}e^{-\alpha_{p}L}$.

\subsection{Propagation Losses ($\alpha$) and Transmittances ($T$)}

Consider an optical wave with initial power $P_{0}$ that completes a round trip $L = 2\pi r$ in a ring resonator with radius $r$, experiencing attenuation due to a loss $\alpha$ per meter. The relation between these parameters is given by

\begin{equation}
    P(L) = P_{0}e^{-\alpha L} \approx P_{0}(1-\alpha L) \quad \Rightarrow \quad \Delta P = -\alpha L P_{0} \quad \quad \quad (\alpha L \ll 1)
\end{equation}

The round trip time $\Delta t$ for the optical signal, which moves with a group velocity $v_{g}$ within the ring resonator, can be calculated as

\begin{equation}
    \Delta t = \frac{L}{v_{g}} \quad \Rightarrow \quad \frac{\Delta P}{\Delta t} = -\alpha v_{g} P_{0}
\end{equation}

The term $\alpha v_{g}$ represents the intrinsic coupling rate $\kappa_{i}$. Thus, the propagation losses $\alpha_{p}$ and $\alpha_{s}$ are given by

\begin{equation}\label{alpha_kappa_vg}
    \alpha_{p} = \frac{\kappa_{i,p}}{v_{g,p}} = \frac{1}{2\pi r}\frac{\kappa_{i,p}}{\text{FSR}_{p}} \quad \text{and} \quad \alpha_{s} = \frac{\kappa_{i,s}}{v_{g,s}} = \frac{1}{2\pi r}\frac{\kappa_{i,s}}{\text{FSR}_{s}}
\end{equation}

where FSR$_{p}$ and FSR$_{s}$ are the free spectral ranges (in Hertz) of our ring resonator. To determine $T$ (pump or probe transmittance), we measure the optical power before and after the sample. The power before the sample is $P_{\text{in}}(t)$, and after the sample is $P_{\text{out}}(t) = T^{2}P_{\text{in}}(t)e^{-\alpha L}$, as shown in \Cref{Pump_powers} and \Cref{Probe_powers}. Using a power meter, we measure the DC components of these optical powers to find $T_{p}$ and $T_{s}$ as

\begin{equation}
    T_{p} = \sqrt{\frac{P_{p,\text{out}}^{\text{DC}}}{P_{p,\text{in}}^{\text{DC}}}}e^{\alpha_{p}L/2} \quad \text{and} \quad T_{s} = \sqrt{\frac{P_{s,\text{out}}^{\text{DC}}}{P_{s,\text{in}}^{\text{DC}}}}e^{\alpha_{s}L/2}
\end{equation}

As the pump and probe in this work are only TE or TM, we can improve the notation: instead of $\alpha_{p}$, FSR$_{p}$, $\alpha_{s}$, FSR$_{s}$ and so on, we denote them as $\alpha^{\text{TE}}$, FSR$^{\text{TE}}$, $\alpha^{\text{TM}}$ and FSR$^{\text{TM}}$. In \Cref{fig:Fig4_Supp} we show the relation between $\left(P_{p,\text{in}},P_{p,\text{out}}\right)$ and $\left(P_{s,\text{in}},P_{s,\text{out}}\right)$, which can be used to find $T^{\text{TE}}$ and $T^{\text{TM}}$. The deviation from a straight line in \Cref{fig:Fig4_Supp}b suggests a pump reflection due to the low probe power and the high pump powers. This does not reflect the actual Brillouin experiment conditions. The values for $T^{\text{TE}}$ and $T^{\text{TM}}$ are based on data along the straight line and can be found in \Cref{tab:experimental_params}.

\begin{figure}[ht]
    \centering
    \includegraphics{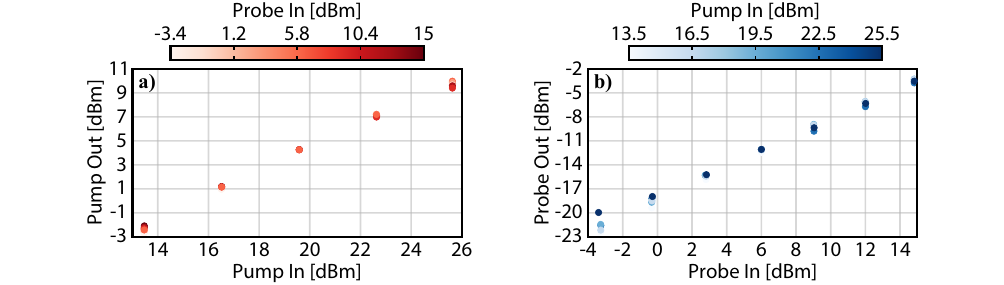}
    \caption{Output vs. input powers for pump (TE, (a)) and probe (TM (b)) lasers, using a waveguide with $\theta=$ \SI{0}{\degree} and width $w=$ \SI{500}{
m}.}
    \label{fig:Fig4_Supp}
\end{figure}

\subsection{Ring Resonator Optical Modes}\label{supp:Ring_Resonator_Optical_Modes}

To understand our device and determine the key parameters for Brillouin gain calculation such as propagation losses, we studied the optical modes of our system. The relevant parameters are shown in \Cref{fig:Fig5_Supp}.

\begin{figure}[ht]
    \centering
    \includegraphics{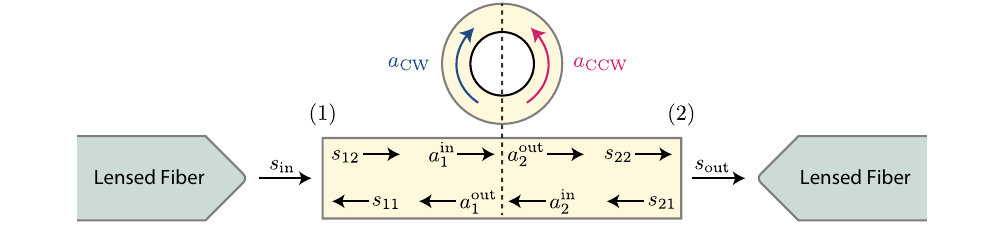}
    \caption{Schematic illustrating the interaction between fiber, waveguide, and ring components, highlighting key elements of the system.}
    \label{fig:Fig5_Supp}
\end{figure}

The interaction between the ring resonator's optical modes and the waveguide is described by the coupled-mode equations given by

\begin{equation}
    \overbrace{\hspace{5pt}s_{11} = a_{1}^{\text{out}}e^{i\delta_{1}}, \quad s_{12} = -s_{11}\sqrt{R_{1}} + s_{\text{in}}\sqrt{T_{1}} \quad \text{and} \quad a_{1}^{\text{in}} = s_{12}e^{i\delta_{1}}\hspace{5pt}}^{\text{equations related to port (1)}}
\end{equation}

\begin{equation}
    \overbrace{\hspace{5pt}a_{1}^{\text{out}} = a_{2}^{\text{in}} - \sqrt{\kappa_{e}}a_{\text{CW}} \quad \text{and} \quad a_{2}^{\text{out}} = a_{1}^{\text{in}} - \sqrt{\kappa_{e}}a_{\text{CCW}}\hspace{5pt}}^{\text{equations interfacing the ring resonator}}
\end{equation}

\begin{equation}
    \overbrace{\hspace{5pt}s_{22} = a_{2}^{\text{out}}e^{i\delta_{2}}, \quad s_{21} = -s_{22}\sqrt{R_{2}}, \quad a_{2}^{\text{in}} = s_{21}e^{i\delta_{2}} \quad \text{and} \quad s_{\text{out}} = s_{22}\sqrt{T_{2}}\hspace{5pt}}^{\text{equations related to port (2)}}
\end{equation}

\begin{align}
    \overbrace{\hspace{5pt}\frac{d}{dt}\left(a_{\text{CCW}}\right) = i\Delta a_{\text{CCW}} - \frac{\kappa}{2}a_\text{CCW} + iJa_{\text{CW}} + \sqrt{\kappa_{e}}a_{1}^{\text{in}}\hspace{5pt}}^{\text{ring resonator mode dynamic equations}}\\
    \frac{d}{dt}\left(a_{\text{CW}}\right) = i\Delta a_{\text{CW}} - \frac{\kappa}{2}a_\text{CW} + iJ^{*}a_{\text{CCW}} + \sqrt{\kappa_{e}}a_{2}^{\text{in}}
\end{align}

Here, $s_{\text{in}}$ is the input field and $s_{\text{out}}$ is the measured output field. The coefficients $R_{1}$ and $R_{2}$ are the reflectances, while $T_{1}$ and $T_{2}$ are the transmittances ($T_{j}$ + $R_{j}$ = 1); $\Delta$ represents the detuning and is defined as $\Delta = \omega_{p} - \omega_{0}$, the difference between the laser frequency $\omega_{p}$ and the cavity resonance frequency $\omega_{0}$; $\delta_{1}$ and $\delta_{2}$ are the phases from port 1 to cavity and from port 2 to cavity, respectively; $\kappa_{e}$ is the coupling between the bus waveguide and the ring (external coupling rate); $\kappa$ is the total coupling rate and $J$ is the coupling strength between the counterpropagating resonator modes. The subscripts ``CCW" and ``CW" stand for ``counterclockwise" and ``clockwise", respectively. To solve this set of coupled differential equations we assume $J \approx 0$ and equal facet reflectance’s $R_{1} \approx R_{2} \approx R$. Consequently, the stationary solution for $a_{\text{CCW}}$ and the transmission coefficient $s_{\text{out}}/s_{\text{in}}$ are given by

\begin{equation}
    a_{\text{CCW}} = \frac{\sqrt{\kappa_{e}}}{\kappa/2 - i\Delta}a_{1}^{\text{in}} \quad \Rightarrow \quad \frac{s_{\text{out}}}{s_{\text{in}}} = \frac{\left(\kappa/2-i\Delta\right)\left(\kappa/2 - \kappa_{e}-i\Delta\right)\left(1-R\right)e^{i\left(\delta_{1}+\delta_{2}\right)}}{\left(\kappa/2 - i\Delta\right)^{2} - \left(\kappa/2 - \kappa_{e} -i\Delta\right)^{2}Re^{2i\left(\delta_{1}+\delta_{2}\right)}}
\end{equation}

Since our photoreceiver maintains flat gain throughout the frequency range of interest, the voltage $V$ measured at the oscilloscope is proportional to the optical power, which, in turn, is proportional to $|s_{\text{out}}/s_{\text{in}}|^{2}$. Therefore, we have

\begin{equation}
    \frac{V}{V_{0}} = \left|\frac{s_{\text{out}}}{s_{\text{in}}}\right|^{2} = \left|\frac{\left(\kappa/2-i\Delta\right)\left(\kappa/2 - \kappa_{e}-i\Delta\right)\left(1-R\right)}{\left(\kappa/2 - i\Delta\right)^{2} - \left(\kappa/2 - \kappa_{e} -i\Delta\right)^{2}Re^{2i\left(\delta_{1}+\delta_{2}\right)}}\right|^{2}
\end{equation}

for some constant $V_{0}$. It's worth noting that this expression recovers two well-known regimes: firstly, when $\kappa_{e} = 0$, effectively implying the absence of the ring structure, resulting in the measurement of just the Fabry-Perot effect; secondly, when $R = 0$, indicating no reflection and yielding the well-known cavity transmission, i.e.,

\begin{equation}
    \frac{V(\kappa_{e} = 0)}{V_{0}} = \frac{\left(1-R\right)^{2}}{1+R^{2}-2R\cos{\left[2\left(\delta_{1}+\delta_{2}\right)\right]}} \quad \quad \text{and} \quad \quad \frac{V(R = 0)}{V_{0}} = 1 - \frac{\kappa_{e}\left(\kappa - \kappa_{e}\right)}{\kappa^{2}/4 + \Delta^{2}}
\end{equation}

The final step necessary to fully utilize the expression for fitting the optical mode is to determine the dependence that $\delta_{1}$ and $\delta_{2}$ have with $\Delta$, i.e., $\delta_{1} = \delta_{1}(\Delta)$ and $\delta_{2} = \delta_{2}(\Delta)$. We will assume a linear dependence with $\Delta$, which allow us to express $\delta_{1} + \delta_{2} = (\delta_{0}/2)\left(\Delta - \Delta_{0}\right)$, and then the final expression is given by

\begin{equation}\label{fitting_full}
    \frac{V(\Delta)}{V_{0}} = \left|\frac{\left(\kappa/2-i\Delta\right)\left(\kappa/2 - \kappa_{e}-i\Delta\right)\left(1-R\right)}{\left(\kappa/2 - i\Delta\right)^{2} - \left(\kappa/2 - \kappa_{e} -i\Delta\right)^{2}Re^{i\delta_{0}\left(\Delta-\Delta_{0}\right)}}\right|^{2}
\end{equation}

The modes of our ring resonator are shown in \Cref{fig:Fig6_Supp}a. We have used the fitting expression outlined in \Cref{fitting_full} to model these modes, represented by the black dashed curves in \Cref{fig:Fig6_Supp}(b,c). The agreement between the modeled modes and the experimental data is excellent. If the resonances are split into CCW and CW modes, the coupling coefficient $J$ must be incorporated into the algebra. Additionally, when allowing $R_{1} \neq R_{2}$, the transmission coefficient can be expressed as follows

\begin{equation}
    \frac{s_{\text{out}}}{s_{\text{in}}} = \frac{\left[\left(\kappa/2-i\Delta\right)\left(\kappa/2 - \kappa_{e}-i\Delta\right) +J^{2}\right]\sqrt{\left(1-R_{1}\right)\left(1-R_{2}\right)}e^{i\left(\delta_{1}+\delta_{2}\right)}}{\left(\kappa/2 - i\Delta\right)^{2} +J^{2} - \left[\left(\kappa/2 - \kappa_{e}-i\Delta\right)^{2} + J^{2}\right]\sqrt{R_{1}R_{2}}e^{2i\left(\delta_{1}+\delta_{2}\right)} - i\kappa_{e}J\left(\sqrt{R_{1}}e^{2i\delta_{1}} + \sqrt{R_{2}}e^{2i\delta_{2}}\right)}
\end{equation}

\begin{figure}[ht]
    \centering
    \includegraphics{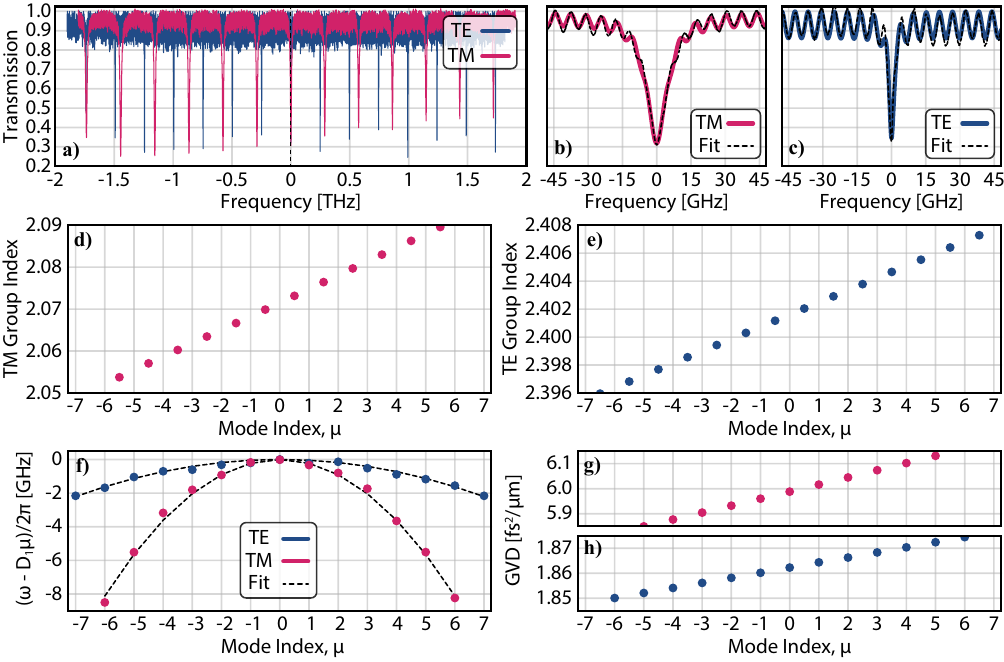}
    \caption{\textbf{(a)} Experimental optical spectrum of the $Z$-cut microring resonator of \SI{80}{\um} radius, displaying both TE and TM mode families. The modes $\mu=0$ from both TE and TM families are aligned at the center of the plot. Relative frequency measurements were performed using a Mach-Zehnder Interferometer (MZI) with a 137 MHz free spectral range (FSR), calibrated using an HCN gas cell; \textbf{(b,c)} Curve fitting for the $\mu = 0$ optical modes of TM and TE, respectively, illustrated by the black dashed lines. The derived fit parameters are detailed in \Cref{tab:experimental_params}; \textbf{(d,e)} Group index of TE and TM modes, respectively; \textbf{(f)} Dispersion analysis of both TE and TM families; \textbf{(g,h)} Values of the group velocity dispersion (GVD) for both TE and TM families. We used the waveguide with $\theta=$ \SI{60}{\degree} and $w=$ \SI{500}{\nm} for these measurements.}
    \label{fig:Fig6_Supp}
\end{figure}

The group index for each mode is shown in \Cref{fig:Fig6_Supp}(d,e) for TE and TM families. The mode index for each $n_{g}$ is a half-integer, determined by finite differences between frequencies. One data point is lost; thus $n_{g}$ is assigned to the midpoint between modes. The mode dispersion is in \Cref{fig:Fig6_Supp}f, fitted using the second-order Taylor expansion for the angular frequency as

\begin{equation}
    \omega = \omega_{0} + D_{1}\mu + \frac{D_{2}}{2}\mu^{2}
\end{equation}

show normal dispersion for both the TE and TM families. The group velocity dispersion (GVD) values are shown in \Cref{fig:Fig6_Supp}(g,h). Additional parameters of \Cref{fig:Fig6_Supp} are listed in \Cref{tab:experimental_params} with other experimental parameters.

\newpage

\subsection{Optical Fiber Brillouin Gain ($G_{B}^{\text{fiber}}$)}

In principle, we could determine $G_{B}^{\text{fiber}}$ simply by referring to (and relying on) the typical values for SMF28. However, the Brillouin gain varies significantly with germanium content. Studies show variations from 0.3 W$^{-1}$m$^{-1}$ to 1.0 W$^{-1}$m$^{-1}$ for GeO$_{2}$ core content of 3\% to 10\%~\cite{633570}. Therefore, it is often good practice to measure $G_{B}^{\text{fiber}}$ for your specific setup. Theoretical Brillouin gain for pure silica, $G_{B}^{\text{silica}}$, can be calculated analytically~\cite{PhysRevA.92.013836,10.1063/1.5088169}, as shown in \Cref{GB_silica} using values from \Cref{tab:silica_typical_params}.

\begin{equation}\label{GB_silica}
    G_{B}^{\text{silica}} = \frac{1}{A_{\text{eff}}}\frac{4\pi^{2} n^{7}p_{12}^{2}}{c\rho \lambda^{2}\Gamma_{B}V_{L}} = 0.26 \text{ [W$^{-1}$m$^{-1}$]}
\end{equation}

\begin{table}[ht]
    \centering
    \caption{Parameters utilized in the computation of Brillouin gain for pure silica.}
    \label{tab:silica_typical_params}
    \begin{tabular}{clcc}
        \toprule
        Symbol & Description & Value & Unit \\
        \midrule
        $c$ & Speed of light & 299792458 & m/s\\
        \addlinespace
        $p_{12}$ & Silica photoelastic coefficient & 0.271 & 1\\
        \addlinespace
        $\lambda$ & Optical wavelength & 1550 & nm\\
        \addlinespace
        $n$ & Silica refractive index at 1550 nm & 1.44 & 1\\
        \addlinespace
        $\rho$ & Silica density & 2203 & kg/m$^{3}$\\
        \addlinespace
        $V_{L}$ & Silica longitudinal velocity & 5969 & m/s\\
        \addlinespace
        $\Gamma_{B}/2\pi$ & Acoustic linewidth & 30 & MHz\\
        \addlinespace
        $A_{\text{eff}}$ & Effective optical mode area & 80 & $\mu$m$^{2}$\\
        \bottomrule
    \end{tabular}
\end{table}

In our setup, we use fiber pigtails from different manufacturers, and although they are SMF28 compatible fibers, we can only specify that these systems combined have an effective Brillouin shift of \SI{10.82}{\GHz}, as shown in \Cref{fig:Fig7_Supp}(a,b). To calculate $G_{B}^{\text{fiber}}$, we removed the sample and connected the lensed fibers face to face. This scenario can be analyzed using \Cref{Pout_silica} with $L = 0$ and $T_{p}^{2} \rightarrow T$, $T_{s}^{2} \rightarrow T$, since only a single transmittance is required in the absence of sample interfaces. The lens fibers are of different brands, which increases mismatching and losses, and we measured $T=0.5$ in our system. The output power and gain are shown in \Cref{Pout_silica_without_chip} as

\begin{equation}\label{Pout_silica_without_chip}
    P_{s,\text{out}}^{\text{fiber}} = TP_{s,\text{in}}e^{G_{B}^{\text{fiber}}P_{p,\text{in}}\left(L_{p} + TL_{s}\right)} \quad \Rightarrow \quad
    G_{B}^{\text{fiber}} = \frac{1}{P_{p,\text{in}}}\frac{1}{\left(L_{p}+TL_{s}\right)}\ln{\left(\frac{1}{T}\frac{P_{s,\text{out}}^{\text{fiber}}}{P_{s,\text{in}}}\right)}
\end{equation}

Since the lasers are not modulated here, $P_{p,\text{in}}$ and $P_{s,\text{in}}$ are constant. Sweeping the probe wavelength around the Brillouin region with a piezoelectric fine scanning tuning, we measure the AC part of the optical power output, $\delta P_{s,\text{out}}^{\text{fiber}}$, as follows

\begin{equation}
    P_{s,\text{out}}^{\text{fiber}} = TP_{s,\text{in}} + \delta P_{s,\text{out}}^{\text{fiber}} \quad \Rightarrow \quad G_{B}^{\text{fiber}} = \frac{1}{P_{p,\text{in}}}\frac{1}{\left(L_{p}+TL_{s}\right)}\ln{\left(1 + \frac{1}{T}\frac{\delta P_{s,\text{out}}^{\text{fiber}}}{P_{s,\text{in}}}\right)}
\end{equation}

\begin{figure}[ht]
    \centering
    \includegraphics{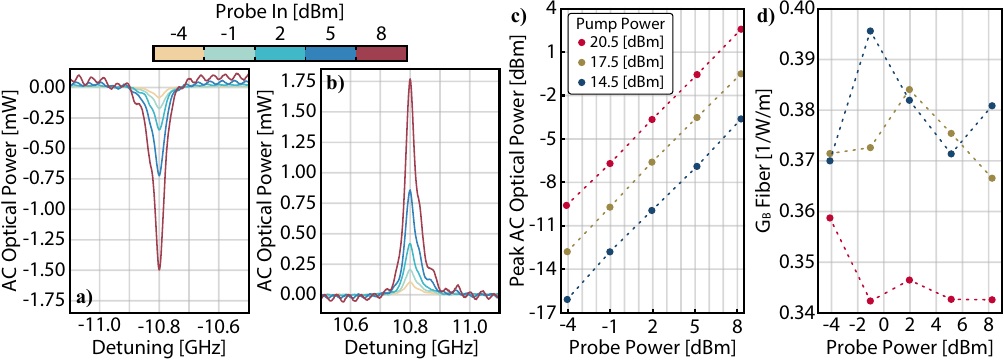}
    \caption{\textbf{(a,b)} Anti-Stokes and Stokes signals at pump power of 20.5 [dBm]; \textbf{(c)} Peak Stokes signal as a function of the probe power, showing the linear dependency between them as expected in \Cref{Pout_silica_without_chip}; \textbf{(d)} Computed Brillouin gain for the fiber for all the points shown in \Cref{fig:Fig7_Supp}c.}
    \label{fig:Fig7_Supp}
\end{figure}

In \Cref{fig:Fig7_Supp}c we have the fiber SBS gain as a function of the input probe power. In \Cref{fig:Fig7_Supp}d we show the values of $G_{B}^{\text{fiber}}$, in which we choose a conservative estimate of $G_{B}^{\text{fiber}} =$ \SI{0.36}{\meter^{-1}\watt^{-1}} for our calculations.

\newpage

\subsection{Previous Sample}

The initial version of this article was uploaded to arXiv using a different sample that did not include waveguides with varying angles. This limitation restricted our discussion on mechanical angular dispersion. However, we successfully observed cross-polarized stimulated Brillouin scattering in this sample, as shown in \Cref{fig:Fig8_Supp}(a,b). Additionally, our preliminary simulations, reproduced in \Cref{fig:Fig8_Supp}(d,e,f) with the assumption of a fixed waveguide base, also confirmed the presence of strong inter-polarization scattering in lithium niobate waveguides, even without employing the full model that incorporates piezoelectricity.

\begin{figure}[ht]
    \centering
    \includegraphics{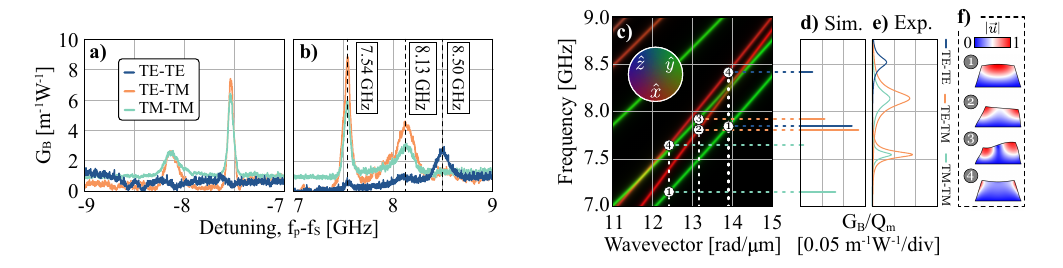}
    \caption{\textbf{(a,b)} Anti-Stokes and Stokes spectrum; \textbf{(c)} Mechanical dispersion diagram for a bottom clamped LN waveguide oriented along the $Y$-axis; \textbf{(d,e)} Simulated and experimental normalized Brillouin gain ($G_{B}/Q_{m}$) for the different pump/probe polarization choices, respectively; \textbf{(f)} Mechanical mode profiles of the modes labeled in \textbf{(a)}.} 
    \label{fig:Fig8_Supp}
\end{figure}

\begin{table}[h!]
    \centering
    \caption{Summary of the parameters used along the article.}
    \label{tab:experimental_params}
    \begin{tabular}{clcc}
        \toprule
        Symbol & Description & Value & Unit \\
        \midrule
        $L_{p}$ & Fiber length from $P_{p,\text{in}}$ to $P_{s,1}$ & 6 & m\\
        \addlinespace
        $L_{s}$ & Fiber length from $P_{s,\text{in}}$ to $P_{p,3}$ & 9.8 & m\\
        \addlinespace
        $r$ & Ring radius & 80 & $\mu$m\\
        \addlinespace
        $t$ & Waveguide thickness & 400 & nm\\
        \addlinespace
        $\delta t$ & Slab thickness & 30 & nm\\
        \addlinespace
        $\phi$ & Wedge angle & 65 & degree\\
        \addlinespace
        $f_{p}$ & Pump modulation frequency & 3.00 & MHz\\
        \addlinespace
        $f_{s}$ & Probe modulation frequency & 3.07 & MHz\\
        \addlinespace
        $\delta f = f_{s} - f_{p}$ & LIA demodulation frequency & 70 & kHz\\
        \addlinespace
        $D_{1}^{\text{TE}}/2\pi$ and $D_{1}^{\text{TM}}/2\pi$ & TE and TM free spectral ranges of the ring resonator & 248.34 and 287.91 & GHz\\
        \addlinespace
        $D_{2}^{\text{TE}}/2\pi$ and $D_{2}^{\text{TM}}/2\pi$ & TE and TM dispersions of the ring resonator & -90.08 and -451.34 & MHz\\
        \addlinespace
        $n_{g}^{\text{TE}}$ and $n_{g}^{\text{TM}}$ & TE and TM group index of the ring resonator & 2.402 and 2.072 & 1\\
        \addlinespace
        $\alpha^{\text{TE}}$ and $\alpha^{\text{TM}}$ & Propagation losses of the ring TE and TM modes & 25.78 and 96.44 & m$^{-1}$\\
        \addlinespace
        $\alpha^{\text{TE}}_{\text{dB}}$ and $\alpha^{\text{TM}}_{\text{dB}}$ & Propagation losses of the ring TE and TM modes in dB & 1.12 and 4.19 & dB/cm\\
        \addlinespace
        $v_{g}^{\text{TE}}$ and $v_{g}^{\text{TM}}$ & TE and TM group velocities of the ring resonator & 1.25$\times$10$^{8}$ and 1.45$\times$10$^{8}$ & m/s\\
        \addlinespace
        $\kappa_{e}^{\text{TE}}/2\pi$ and $\kappa_{e}^{\text{TM}}/2\pi$ & External TE and TM coupling rates & 2.20 and 8.45 & GHz\\
        \addlinespace
        $\kappa_{i}^{\text{TE}}/2\pi$ and $\kappa_{i}^{\text{TM}}/2\pi$ & Intrinsic TE and TM coupling losses & 0.51 and 2.22 & GHz\\
        \addlinespace
        $Q_{i}^{\text{TE}}$ and $Q_{i}^{\text{TM}}$ & Intrinsic TE and TM quality factors & 3.8$\times$10$^{5}$ and 8.7$\times$10$^{4}$ & 1\\
        \addlinespace
        $\Delta_{0}^{\text{TE}}$ and $\Delta_{0}^{\text{TM}}$ & Detunings in \Cref{fig:Fig6_Supp}(b,c) & 3.76 and 3.20 & GHz\\
        \addlinespace
        $\delta_{0}^{\text{TE}}$ and $\delta_{0}^{\text{TM}}$ & Fabry-Perot period in \Cref{fig:Fig6_Supp}(b,c) & 174.07 and 147.10 & ps\\
        \addlinespace
        $R^{\text{TE}}$ and $R^{\text{TM}}$ & Reflectance’s of \Cref{fig:Fig6_Supp}(b,c) & 3.72 and 1.76 & \%\\
        \addlinespace
        $T$ & Fiber-to-fiber transmittance & 0.50 & 1\\
        \addlinespace
        $T^{\text{TE}}$ & Fiber-to-sample TE mode transmittance & 0.19 & 1\\
        \addlinespace
        $T^{\text{TM}}$ & Fiber-to-sample TM mode transmittance & 0.19 & 1\\
        \addlinespace
        $G_{B}^{\text{fiber}}$ & Silica fiber Brillouin gain & 0.36 & m$^{-1}$W$^{-1}$\\
        \bottomrule
    \end{tabular}
\end{table}

\newpage

\begin{table}[ht]
    \centering
    \caption{Partitioning the waveguide size by segments, ensuring precision in total length within a maximum error of \SI{100}{\um}.}
    \label{tab:L2_values}
    \begin{tabular}{ccccc}
        \toprule
        Angle (degree) & Straight (\si{\micro\metre}) & Middle (\si{\micro\metre}) & Curved (\si{\micro\metre}) & Total Length, $L$ (\si{\micro\metre}) \\
        \midrule
        -60 & 6200 & 5000 & 209 & 11409 \\
        -55 & 5848 & 5000 & 192 & 11040 \\
        -50 & 5544 & 5000 & 174 & 10718 \\
        -45 & 5337 & 5000 & 157 & 10494 \\
        -40 & 5125 & 5000 & 140 & 10265 \\
        -35 & 4906 & 5000 & 122 & 10028 \\
        -30 & 4750 & 5000 & 105 & 9855 \\
        -25 & 4595 & 5000 & 87 & 9682 \\
        -20 & 4560 & 5000 & 70 & 9630 \\
        -15 & 3842 & 5000 & 52 & 8894 \\
        -10 & 3818 & 5000 & 35 & 8853 \\
        -5 & 3893 & 5000 & 17 & 8910 \\
        0 & 3800 & 5000 & 0 & 8800 \\
        5 & 3693 & 5000 & 17 & 8710 \\
        10 & 3778 & 5000 & 35 & 8813 \\
        15 & 3842 & 5000 & 52 & 8894 \\
        20 & 3970 & 5000 & 70 & 9040 \\
        25 & 4025 & 5000 & 87 & 9112 \\
        30 & 4280 & 5000 & 105 & 9385 \\
        35 & 4506 & 5000 & 122 & 9628 \\
        40 & 4725 & 5000 & 140 & 9865 \\
        45 & 5037 & 5000 & 157 & 10194 \\
        50 & 5344 & 5000 & 174 & 10518 \\
        55 & 5648 & 5000 & 192 & 10840 \\
        60 & 6100 & 5000 & 209 & 11309 \\
        \bottomrule
    \end{tabular}
\end{table}

\end{widetext}

\end{document}